\documentclass[aps,prb,twocolumn,showpacs]{revtex4}

\usepackage{graphicx}
\usepackage{amsmath}
\usepackage{amssymb}

\newcommand{\ou}[2]{#1_{#2\,\uparrow}}
\newcommand{\oud}[2]{#1_{#2\,\uparrow}^\dagger}

\newcommand{\od}[2]{#1_{#2\,\downarrow}}
\newcommand{\odd}[2]{#1_{#2\,\downarrow}^\dagger}

\newcommand{\Sm}[1]{S^{-}_{#1}}
\newcommand{\Sp}[1]{S^{+}_{#1}}

\newcommand{\vv}[1]{{\bf{#1}}}

\newcommand{\Ek}{E_{{\rm{h}}\vv{k}}}
\newcommand{\Ekq}{E_{{\rm{h}}\vv{k+q}}}

\newcommand{\xik}{\bar{\xi}_{\vv{k}}}
\newcommand{\xikq}{\bar{\xi}_{\vv{k+q}}}

\newcommand{\Dk}{\bar{\Delta}_{{\rm{hZ}}}(\vv{k})}
\newcommand{\Dkq}{\bar{\Delta}_{{\rm{hZ}}}(\vv{k+q})}

\newcommand{\nF}{n_{\rm{F}}}

\begin{document}

\title{Doping dependence of Meissner effect in cuprate
superconductors}

\author{Shiping Feng$^{*}$, Zheyu Huang, and Huaisong Zhao}
\affiliation{Department of Physics, Beijing Normal University,
Beijing 100875, China}

\begin{abstract}
Within the $t$-$t'$-$J$ model, the doping dependence of the Meissner
effect in cuprate superconductors is studied based on the kinetic
energy driven superconducting mechanism. Following the linear
response theory, it is shown that the electromagnetic response
consists of two parts, the diamagnetic current and the paramagnetic
current, which exactly cancels the diamagnetic term in the normal
state, and then the Meissner effect is obtained for all the
temperature $T\leq T_{c}$ throughout the superconducting dome. By
considering the two-dimensional geometry of cuprate superconductors
within the specular reflection model, the main features of the
doping and temperature dependence of the local magnetic field
profile, the magnetic field penetration depth, and the superfluid
density observed on cuprate superconductors are well reproduced. In
particular, it is shown that in analogy to the domelike shape of the
doping dependent superconducting transition temperature, the maximal
superfluid density occurs around the critical doping $\delta\approx
0.195$, and then decreases in both lower doped and higher doped
regimes.
\end{abstract}

\pacs{74.25.Ha, 74.25.Nf, 74.20.Mn}

\maketitle

\section{Introduction}

One of the characteristic features of superconductors is the
so-called Meissner effect \cite{schrieffer83}, i.e., a
superconductor is placed in an external magnetic field $B$ smaller
than the upper critical field $B_{c}$, the magnetic field $B$
penetrates only to a penetration depth $\lambda$ (few hundred nm for
cuprate superconductors at zero-temperature) and is excluded from
the main body of the system. This magnetic field penetration depth
is a fundamental parameter of superconductors, and is closely
related to the superfluid density $\rho_{s}$
\cite{schrieffer83,bonn96}, which is proportional to the squared
amplitude of the macroscopic wave function, and therefore describes
the superconducting (SC) charge carriers. Furthermore, the variation
of the magnetic field penetration depth (then the superfluid
density) as a function of doping and temperature provides
information crucial to understanding the details of the SC state
\cite{schrieffer83,bonn96}. In particular, since the compounds of
cuprate superconductors are doped Mott insulators with the strong
short-range antiferromagnetic correlation dominating the entire SC
phase \cite{damascelli03}, the magnetic field can be also used to
probe the doping and momentum dependence of the SC gap and spin
structure of the Cooper pair \cite{bonn96}. This is why the first
evidence of the d-wave Cooper pairing state in cuprate
superconductors was obtained from the earlier experimental
measurement for the magnetic field penetration depth \cite{hardy93}.
Since then this d-wave SC state remains one of the cornerstones of
our understanding of the physics in cuprate superconductors
\cite{damascelli03,tsuei00}.

Experimentally, by virtue of systematic studies using the
muon-spin-rotation measurement technique, some essential features of
the evolution of the magnetic field penetration depth and superfluid
density in cuprate superconductors with doping and temperature have
been established now for all the temperature $T\leq T_{c}$
throughout the SC dome: (1) the magnetic field screening is found to
be of exponential character \cite{khasanov04,suter04}, in support of
a local (London-type) nature of the electrodynamic response
\cite{schrieffer83}; (2) the magnetic field penetration depth is a
linear temperature dependence at low temperatures except for the
extremely low temperatures where a strong deviation from the linear
characteristics (a nonlinear effect) appears
\cite{hardy93,kamal98,jackson00,pereg07}; (3) the doping dependence
of the superfluid density is observed
\cite{niedermayer93,bernhard01,broun07,khasanov10}, where the
superfluid density exhibits a peak around the critical doping
$\delta\approx 0.19$, and then decreases at both lower doped and
higher doped regimes. This in turn gives rise to the domelike shape
of the doping dependence of the SC transition temperature. In
particular, it has been shown experimentally \cite{bernhard01} that
the peak of the superfluid density at the critical doping
$\delta\approx 0.19$ is a common feature of the hole-doped cuprate
superconductors. Theoretically, to the best of our knowledge, all
theoretical calculations of the experimental measurements of the
magnetic field penetration depth and the related superfluid density
in cuprate superconductors performed so far are based on the
phenomenological d-wave Bardeen-Cooper-Schrieffer (BCS) formalism
\cite{yip92,kosztin97,franz97,li00,sheehy04}. In the local limit,
where the magnetic field penetration depth $\lambda$ is much larger
than the coherence length $\zeta$, i.e., $\lambda\gg\zeta$, it has
been shown \cite{tsuei00,kosztin97} that the simple d-wave pairing
state gives the linear temperature dependence of the magnetic field
penetration depth $\Delta\lambda(T)=\lambda(T)-\lambda(0)\propto
T/\Delta_{0}$ at low temperatures, with $\Delta_{0}$ is the
zero-temperature value of the d-wave gap amplitude. However, the
characteristic feature of the d-wave energy gap is the existence of
the gap nodes on the Fermi surface, which can lead to the nonlinear
effect of field on the penetration depth (then superfluid density)
at the extremely low temperatures, associated with a field induced
increase in the density of available excitation states located
around the gap nodes on the Fermi surface
\cite{yip92,kosztin97,franz97,li00,sheehy04}. This follows from a
fact that the nonlocal effect is closely related to the divergence
of the coherence length $\zeta$ at the gap nodes on the Fermi
surface, as the coherence length $\zeta$ varies in inverse
proportion to the value of the energy gap, then this nonlocal effect
at the extremely low temperatures can lead to a nonlinear
temperature dependence of the magnetic field penetration depth in
the clean limit \cite{yip92,kosztin97,franz97,li00,sheehy04}.

In our recent work \cite{krzyzosiak10} based on the nearest
neighbors hopping $t$-$J$ model, the weak electromagnetic response
in cuprate superconductors has been discussed within the kinetic
energy driven SC mechanism \cite{feng0306}. However, it has been
shown experimentally \cite{damascelli03,well98} that although the
highest energy filled electron band is well described by the nearest
neighbors hopping $t$-$J$ model, the overall dispersion may be
properly accounted by generalizing the nearest neighbors hopping
$t$-$J$ model to include the second- and third-nearest neighbors
hopping terms $t'$ and $t''$. Furthermore, the experimental analysis
\cite{tanaka04} showed that the SC transition temperature (then the
superfluid density) for different families of cuprate
superconductors is strongly correlated with $t'$. In this paper
based on the $t$-$t'$-$J$ model, we study the doping dependence of
the Meissner effect in cuprate superconductors for all the
temperature $T\leq T_{c}$ throughout the SC dome, where one of our
main results is that the superfluid density increases with
increasing doping in the lower doped regime, and reaches a maximum
around the critical doping $\delta\approx 0.195$, then decreases in
the higher doped regime.

The rest of this paper is organized as follows. The basic formalism
is presented in Section \ref{emressect}, where within the
$t$-$t'$-$J$ model, we evaluate the diamagnetic and paramagnetic
components of the response kernel function based on the kinetic
energy driven SC mechanism, and then obtain explicitly the Meissner
effect in cuprate superconductors for all the temperature $T\leq
T_{c}$ throughout the SC dome. Within this theoretical framework, we
discuss the basic behavior of cuprate superconductors in a weak
electromagnetic field in Section \ref{meissner}, and qualitatively
reproduce all the main features of the doping and temperature
dependence of the local magnetic field profile, the magnetic field
penetration depth, and the superfluid density. Finally, we give a
summary and discussions in Section \ref{conclusions}. In Appendix
\ref{limit1} we give the explicit forms of the paramagnetic part of
the response kernel function at the temperatures $T=0$ and
$T=T_{c}$, respectively.

\section{Theoretical framework}
\label{emressect}

In cuprate superconductors, the characteristic feature is the
presence of the CuO$_{2}$ plane \cite{damascelli03}. It has been
argued that the essential physics of the doped CuO$_{2}$ plane
\cite{anderson87,damascelli03} is properly accounted by the
$t$-$t'$-$J$ model on a square lattice. However, for discussions of
the doping and temperature dependence of the Meissner effect in
cuprate superconductors, the $t$-$t'$-$J$ model can be extended by
including the exponential Peierls factors as,
%\begin{widetext}
\begin{eqnarray}\label{tjham}
H&=&-t\sum_{l\hat{\eta}\sigma}e^{-i({e}/{\hbar})\vv{A}(l)\cdot
\hat{\eta}}C^{\dagger}_{l\sigma}C_{l+\hat{\eta}\sigma}+\mu\sum_{l\sigma}
C^{\dagger}_{l\sigma}C_{l\sigma}\nonumber\\
&+&t'\sum_{l\hat{\eta}'\sigma}e^{-i({e}/{\hbar})\vv{A}(l)\cdot
\hat{\eta}'}C^{\dagger}_{l\sigma}C_{l+\hat{\eta}'\sigma}
+J\sum_{l\hat{\eta}}{\bf S}_{l}\cdot {\bf S}_{l+\hat{\eta}},~~~~
\end{eqnarray}
%\end{widetext}
supplemented by an important on-site local constraint $\sum_{\sigma}
C^{\dagger}_{l\sigma}C_{l\sigma}\leq 1$ to remove the double
occupancy, where $\hat{\eta}=\pm\hat{x},\pm\hat{y}$, $\hat{\eta}'
=\pm\hat{x}\pm\hat{y}$, $C^{\dagger}_{l\sigma}$ ($C_{l\sigma}$) is
the electron creation (annihilation) operator, ${\bf S}_{l}=
(S^{x}_{l},S^{y}_{l}, S^{z}_{l})$ are spin operators, and $\mu$ is
the chemical potential. The exponential Peierls factors account for
the coupling of electrons to the weak external magnetic field
\cite{hirsch92,misawa94} in terms of the vector potential
$\vv{A}(l)$. To incorporate the electron single occupancy local
constraint, the charge-spin separation (CSS) fermion-spin theory
\cite{feng0304} has been proposed, where the physics of no double
occupancy is taken into account by representing the electron as a
composite object created by $C_{l\uparrow}= h^{\dagger}_{l\uparrow}
S^{-}_{l}$ and $C_{l\downarrow}=h^{\dagger}_{l\downarrow}
S^{+}_{l}$, with the spinful fermion operator $h_{l\sigma}=
e^{-i\Phi_{l\sigma}}h_{l}$ that describes the charge degree of
freedom of the electron together with some effects of spin
configuration rearrangements due to the presence of the doped hole
itself (charge carrier), while the spin operator $S_{l}$ represents
the spin degree of freedom of the electron, then the electron single
occupancy local constraint is satisfied in analytical calculations.
In particular, it has been shown that under the {\it decoupling
scheme}, this CSS fermion-spin representation is a natural
representation of the constrained electron defined in the Hilbert
subspace without double electron occupancy \cite{feng07}. In this
CSS fermion-spin representation, the $t$-$t'$-$J$ model
(\ref{tjham}) can be expressed as,
%\begin{widetext}
\begin{eqnarray}\label{cssham}
H&=&t\sum_{l\hat{\eta}}e^{-i({e}/{\hbar})\vv{A}(l)\cdot\hat{\eta}}
(h^{\dagger}_{l+\hat{\eta}\uparrow}h_{l\uparrow}S^{+}_{l}
S^{-}_{l+\hat{\eta}}\nonumber\\
&+&h^{\dagger}_{l+\hat{\eta}\downarrow}
h_{l\downarrow}S^{-}_{l}S^{+}_{l+\hat{\eta}})\nonumber\\
&-&t'\sum_{l\hat{\eta}'}e^{-i({e}/{\hbar})\vv{A}(l)\cdot\hat{\eta}'}
(h^{\dagger}_{l+\hat{\eta}'\uparrow}h_{l\uparrow}S^{+}_{l}
S^{-}_{l+\hat{\eta}'}\nonumber\\
&+&h^{\dagger}_{l+\hat{\eta}'\downarrow}
h_{l\downarrow}S^{-}_{l}S^{+}_{l+\hat{\eta}'})\nonumber\\
&-&\mu\sum_{l\sigma} h^{\dagger}_{l\sigma}h_{l\sigma}+J_{{\rm eff}}
\sum_{l\hat{\eta}}{\bf S}_{l}\cdot {\bf S}_{l+\hat{\eta}},
\end{eqnarray}
%\end{widetext}
where $J_{{\rm eff}}=(1-\delta)^{2}J$, and $\delta=\langle
h^{\dagger}_{l\sigma}h_{l\sigma}\rangle=\langle h^{\dagger}_{l}
h_{l}\rangle$ is the charge carrier doping concentration.

As in the conventional superconductors, the key phenomenon occurring
in cuprate superconductors in the SC state is the pairing of charge
carriers \cite{damascelli03,tsuei00}. The system of charge carriers
forms pairs of bound charge carriers in the SC state, while the
pairing means that there is an attraction between charge carriers.
What is the origin of such an attractive force? Recently, the
kinetic energy driven SC mechanism has been developed based on the
$t$-$J$ model \cite{feng0306}, where the charge carrier-spin
interaction from the kinetic energy term in the $t$-$J$ model
(\ref{cssham}) induces a d-wave charge carrier pairing state by
exchanging spin excitations in the higher power of the doping
concentration, then the SC transition temperature is identical to
the charge carrier pair transition temperature. Furthermore, this SC
state is the conventional BCS-like with the d-wave symmetry
\cite{feng07,guo07}, so that the basic d-wave BCS formalism is still
valid in quantitatively reproducing all main low energy features of
the SC coherence of the quasiparticle peaks in cuprate
superconductors, although the pairing mechanism is driven by the
kinetic energy by exchanging spin excitations. Following these
previous discussions \cite{feng0306,feng07,guo07}, the full charge
carrier Green's function in the zero magnetic field case can be
obtained explicitly in the Nambu representation as,
\begin{eqnarray}
\mathbb{G}({\bf{k}},i\omega_n)=Z_{\rm{hF}}\,\frac{i\omega_n\tau_0 +
\bar{\xi}_{\bf{k}}\tau_3 - \bar{\Delta}_{\rm{hZ}}({\bf{k}})
\tau_1}{(i\omega_n)^2 - E_{{\rm{h}}{\bf{k}}}^2},
\label{holegreenfunction}
\end{eqnarray}
where $\tau_{0}$ is the unit matrix, $\tau_{1}$ and $\tau_{3}$ are
Pauli matrices, the renormalized charge carrier excitation spectrum
$\bar{\xi}_{{\bf k}} =Z_{\rm hF}\xi_{\bf k}$, with the mean-field
charge carrier excitation spectrum $\xi_{{\bf k}}=Zt\chi_{1}
\gamma_{{\bf k}}- Zt' \chi_{2}\gamma_{{\bf k}}'- \mu$, the spin
correlation functions $\chi_{1}=\langle S_{i}^{+}
S_{i+\hat{\eta}}^{-}\rangle$, $\chi_{2}=\langle S_{i}^{+}
S_{i+\hat{\eta}'}^{-}\rangle$, $\gamma_{{\bf k}}=
(1/Z)\sum_{\hat{\eta}}e^{i{\bf k}\cdot \hat{\eta}}$, $\gamma_{{\bf
k}}'= (1/Z)\sum_{\hat{\eta}'}e^{i{\bf k} \cdot\hat{\eta}'}$, $Z$ is
the number of the nearest neighbor or second-nearest neighbor sites,
the renormalized charge carrier d-wave pair gap $\bar{\Delta}_{\rm
hZ}({\bf k})=Z_{\rm hF} \bar{\Delta}_{\rm h}({\bf k})$, where the
effective charge carrier d-wave pair gap $\bar{\Delta}_{\rm h}({\bf
k})=\bar{\Delta}_{\rm h}({\rm cos} k_{x}-{\rm cos}k_{y})/2$, and the
charge carrier quasiparticle spectrum $E_{{\rm{h}}{\bf k}}=\sqrt
{\bar{\xi}^{2}_{{\bf k}}+ |\bar{\Delta}_{\rm hZ}({\bf k})|^{2}}$,
while the effective charge carrier gap parameter $\bar{\Delta}_{\rm
h}$ and the quasiparticle coherent weight $Z_{\rm hF}$ have been
determined self-consistently along with other equations
\cite{feng07,guo07}, and then all order parameters and chemical
potential have been determined by the self-consistent calculation.

\subsection{Linear response approach in the presence of a weak
external magnetic field}

In cuprate superconductors, an external magnetic field generally
represents a large perturbation on the system, then the induced
field arising from the superconductor cancels this external magnetic
field over most of the system. In this case, the net field acts only
near the surface on a scale of the magnetic field penetration depth,
and then it can be treated as a weak perturbation on the system as a
whole \cite{schrieffer83}. This is why the Meissner effect can be
successfully studied within the linear response approach
\cite{fetter71,fukuyama69}, where the linear response current
density $J_{\mu}$ and the vector potential $A_{\nu}$ are related by
a nonlocal kernel of the response function $K_{\mu\nu}$ as,
\begin{equation}\label{linres}
J_\mu({\bf{q}},\omega)=-\sum\limits_{\nu=1}^3
K_{\mu\nu}({\bf{q}},\omega)A_\nu({\bf{q}},\omega),
\end{equation}
with the Greek indices label the axes of the Cartesian coordinate
system. The kernel of the response function in Eq. (\ref{linres})
plays a crucial role for the discussion of the doping dependence of
the Meissner effect in cuprate superconductors, and can be separated
into two parts as,
\begin{equation}\label{kernel}
K_{\mu\nu}({\bf{q}},\omega) = K^{({\rm{d}})}_{\mu\nu}({\bf{q}},
\omega) + K^{({\rm{p}})}_{\mu\nu}({\bf{q}},\omega),
\end{equation}
where $K^{({\rm{d}})}_{\mu\nu}$ and $K^{({\rm{p}})}_{\mu\nu}$ are
the corresponding diamagnetic and paramagnetic parts, respectively,
and are closely related to the electron current density in the
presence of the vector potential $A_{\nu}$.

The vector potential $\vv{A}$ (then the weak external magnetic field
${\bf B}=\rm{rot}\,\vv{A}$) has been coupled to the electrons, which
are now represented by $C_{l\uparrow}= h^{\dagger}_{l\uparrow}
S^{-}_{l}$ and $C_{l\downarrow}= h^{\dagger}_{l\downarrow}S^{+}_{l}$
in the CSS fermion-spin representation. In this case, the electron
current operator can be obtained in terms of the electron
polarization operator, which is a summation over all the particles
and their positions, and can be expressed explicitly in the CSS
fermion-spin representation as,
\begin{equation}\label{poloper}
{\bf P}=-e\sum\limits_{i\sigma}{\bf R}_{i}C^{\dagger}_{i\sigma}
C_{i\sigma}=e\sum\limits_{i}{\bf R}_{i}h^{\dagger}_{i} h_{i},
\end{equation}
then the electron current operator is obtained by evaluating the
time-derivative of this polarization operator (\ref{poloper}) as
\cite{mahan00},
%\begin{widetext}
\begin{eqnarray}
\vv{j}&=&{\partial\vv{P}\over\partial t}={i\over\hbar}[H,\vv{P}]
\nonumber\\
&=&{iet\over\hbar}\sum\limits_{l{\hat{\eta}}}{\hat{\eta}}
e^{-i({e\over\hbar})\vv{A}(l)\cdot{\hat{\eta}}}\left(\ou{h}{l}
\oud{h}{l+{\hat{\eta}}}\Sp{l}\Sm{l+{\hat{\eta}}}\right .\nonumber\\
&+&\left .\od{h}{l} \odd{h}{l+{\hat{\eta}}}\Sm{l}\Sp{l+{\hat{\eta}}}
\right)\nonumber\\
&-&{iet'\over\hbar}\sum\limits_{l{\hat{\eta}'}}{\hat{\eta}'}
e^{-i({e\over\hbar})\vv{A}(l)\cdot{\hat{\eta}'}}\left(\ou{h}{l}
\oud{h}{l+{\hat{\eta}'}}\Sp{l}\Sm{l+{\hat{\eta}'}}\right . \nonumber\\
&+&\left . \od{h}{l}
\odd{h}{l+{\hat{\eta}'}}\Sm{l}\Sp{l+{\hat{\eta}'}} \right).
\end{eqnarray}
%\end{widetext}
In the linear response approach, this electron current operator is
reduced as $\vv{j}=\vv{j}^{(d)}+\vv{j}^{(p)}$, with the
corresponding diamagnetic (d) and paramagnetic (p) components of the
electron current operator are given by,
\begin{subequations}
\begin{eqnarray}
\vv{j}^{(\rm{d})}&=&{e^{2}t\over\hbar^{2}}\sum\limits_{l{\hat{\eta}}}
{\hat{\eta}}\vv{A}(l)\cdot{\hat{\eta}}\left(\ou{h}{l}
\oud{h}{l+{\hat{\eta}}}\Sp{l}\Sm{l+{\hat{\eta}}}\right .\nonumber\\
&+&\left. \od{h}{l}\odd{h}{l+{\hat{\eta}}}\Sm{l}\Sp{l+{\hat{\eta}}}
\right)\nonumber\\
&-&{e^{2}t'\over\hbar^{2}}\sum\limits_{l{\hat{\eta}'}}{\hat{\eta}'}
\vv{A}(l)\cdot{\hat{\eta}'}\left(\ou{h}{l} \oud{h}{l+{\hat{\eta}'}}
\Sp{l}\Sm{l+{\hat{\eta}'}}\right.\nonumber\\
&+& \left. \od{h}{l} \odd{h}{l+{\hat{\eta}'}}\Sm{l}
\Sp{l+{\hat{\eta}'}}\right),\label{tcurdia}\\
\vv{j}^{(\rm{p})}&=&{iet\over\hbar}\sum\limits_{l{\hat{\eta}}}
{\hat{\eta}}\left(\ou{h}{l} \oud{h}{l+{\hat{\eta}}}\Sp{l}
\Sm{l+{\hat{\eta}}}+ \od{h}{l} \odd{h}{l+{\hat{\eta}}}\Sm{l}
\Sp{l+{\hat{\eta}}}\right)\nonumber\\
&-&{iet'\over\hbar}\sum\limits_{l{\hat{\eta}'}}{\hat{\eta}'}
\left(\ou{h}{l}\oud{h}{l+{\hat{\eta}'}}\Sp{l}\Sm{l+{\hat{\eta}'}}
\right.\nonumber\\
&+&\left. \od{h}{l}
\odd{h}{l+{\hat{\eta}'}}\Sm{l}\Sp{l+{\hat{\eta}'}}\right),
\label{tcurpara9}
\end{eqnarray}
\end{subequations}
respectively. Obviously, the diamagnetic component of the electron
current operator in Eq. (\ref{tcurdia}) is proportional to the
vector potential, and in this case, we can obtain the diamagnetic
part of the response kernel directly as,
\begin{eqnarray}\label{diakernel}
K_{\mu\nu}^{(\rm{d})}(\vv{q},\omega)=-{4e^{2}\over\hbar^{2}}(\chi_{1}
\phi_{1}t-2\chi_{2}\phi_{2}t')\delta_{\mu\nu}={1\over\lambda^{2}_{L}}
\delta_{\mu\nu},
\end{eqnarray}
where the charge carrier particle-hole parameters $\phi_{1}=\langle
h^{\dagger}_{i\sigma} h_{i+\hat{\eta}\sigma}\rangle$ and $\phi_{2}=
\langle h^{\dagger}_{i\sigma}h_{i+\hat{\eta}'\sigma} \rangle$, while
$\lambda^{-2}_{L}=-4e^{2}(\chi_{1}\phi_{1}t- 2\chi_{2} \phi_{2}t')
/\hbar^{2}$ is the London penetration depth, and is doping and
temperature dependent.

However, the paramagnetic part of the response kernel is more
complicated to calculate, since it involves evaluation of the
following electron current-current correlation function
(polarization bubble),
\begin{equation}\label{corP}
P_{\mu\nu}(\vv{q},\tau)=-\langle T_\tau
\{j^{({\rm{p}})}_{\mu}(\vv{q},\tau) j_{\nu}^{({\rm{p}})}(-\vv{q},0)
\}\rangle ,
\end{equation}
then the paramagnetic part of the response kernel
$K_{\mu\nu}^{(\rm{p})}(\vv{q},\omega)$ can be obtained as
$K_{\mu\nu}^{(\rm{p})}(\vv{q},\omega)=P_{\mu\nu}(\vv{q},\omega)$. In
the CSS fermion-spin approach, the paramagnetic component of the
electron current operator in Eq. (\ref{tcurpara9}) can be decoupled
as,
\begin{eqnarray}
\vv{j}^{(\rm{p})}&=&-{ie\chi_{1}t\over\hbar}
\sum\limits_{l{\hat{\eta}}\sigma}{\hat{\eta}}
h^{\dagger}_{l+{\hat{\eta}}\sigma}h_{l\sigma}+{ie\chi_{2}t'\over\hbar}
\sum\limits_{l{\hat{\eta}'}\sigma}{\hat{\eta}'}
h^{\dagger}_{l+{\hat{\eta}'}\sigma}h_{l\sigma}\nonumber\\
&-&{ie\phi_{1}t\over\hbar}\sum\limits_{l{\hat{\eta}}}
{\hat{\eta}}\left(\Sp{l}\Sm{l+{\hat{\eta}}}+\Sm{l}\Sp{l+{\hat{\eta}}}
\right)\nonumber\\
&+&{ie\phi_{2}t'\over\hbar}\sum
\limits_{l{\hat{\eta}'}}{\hat{\eta}'}
\left(\Sp{l}\Sm{l+{\hat{\eta}'}}+\Sm{l}\Sp{l+{\hat{\eta}'}}\right),
\label{tcurpara15}
\end{eqnarray}
where the third and fourth terms in the right-hand side refer to the
contribution from the electron spin, and can be expressed explicitly
as,
\begin{widetext}
\begin{eqnarray*}
&-&{ie\phi_{1}t\over\hbar}\sum\limits_{l\hat{\mu}}
{\hat{\mu}}[(\Sp{l}\Sm{l+{\hat{\mu}}}+\Sm{l}\Sp{l+{\hat{\mu}}})-
(\Sp{l}\Sm{l-{\hat{\mu}}}+\Sm{l}\Sp{l-{\hat{\mu}}})]
={ie\phi_{1}t\over\hbar}\sum\limits_{l\hat{\mu}}
{\hat{\mu}}[(\Sp{l+{\hat{\mu}}}\Sm{l}+\Sm{l+{\hat{\mu}}}\Sp{l})
-(\Sp{l}\Sm{l+{\hat{\mu}}}+\Sm{l}\Sp{l+{\hat{\mu}}})]\\
&\equiv& 0,\\
&~&{ie\phi_{2}t'\over\hbar}\sum\limits_{l}[({\hat{x}}+{\hat{y}})
(\Sp{l}\Sm{l+{\hat{x}}+{\hat{y}}}+\Sm{l}\Sp{l+{\hat{x}}+{\hat{y}}})
-({\hat{x}}+{\hat{y}})(\Sp{l}\Sm{l-{\hat{x}}-{\hat{y}}}+\Sm{l}
\Sp{l-{\hat{x}}-{\hat{y}}})+({\hat{x}}-{\hat{y}})(\Sp{l}
\Sm{l+{\hat{x}}-{\hat{y}}}+\Sm{l}\Sp{l+{\hat{x}}-{\hat{y}}})\\
&-&({\hat{x}}-{\hat{y}})(\Sp{l}\Sm{l-{\hat{x}}+{\hat{y}}}+
\Sm{l}\Sp{l-{\hat{x}}+{\hat{y}}})]\\
&=&{ie\phi_{2}t'\over\hbar}\sum\limits_{l}[({\hat{x}}+{\hat{y}})
(\Sp{l}\Sm{l+{\hat{x}}+{\hat{y}}}+\Sm{l}\Sp{l+{\hat{x}}+{\hat{y}}})
-({\hat{x}}+{\hat{y}})(\Sp{l+{\hat{x}}+{\hat{y}}}\Sm{l}+
\Sm{l+{\hat{x}}+{\hat{y}}}\Sp{l})+({\hat{x}}-{\hat{y}})
(\Sp{l}\Sm{l+{\hat{x}}-{\hat{y}}}+\Sm{l}\Sp{l+{\hat{x}}-{\hat{y}}})\\
&-&({\hat{x}}-{\hat{y}})
(\Sp{l+{\hat{x}}-{\hat{y}}}\Sm{l}+\Sm{l+{\hat{x}}-{\hat{y}}}\Sp{l})]
\equiv 0.
\end{eqnarray*}
\end{widetext}
In this case, the majority contribution for the paramagnetic
component of the electron current operator comes from the electron
charge, and then the paramagnetic component of the electron current
operator in Eq. (\ref{tcurpara15}) can be expressed  explicitly as,
\begin{eqnarray}
j_{\mu}^{(\rm{p})}&=&-{ie\chi_{1}t\over\hbar}\sum\limits_{l\sigma}
(h^{\dagger}_{l+\hat{\mu}\sigma}h_{l\sigma}-h^{\dagger}_{l\sigma}
h_{l+\hat{\mu}\sigma})\nonumber\\
&+&{ie\chi_{2}t'\over\hbar}\sum\limits_{l\sigma\nu\neq\mu}
[(h^{\dagger}_{l+\hat{\mu}+\hat{\nu}\sigma}+
h^{\dagger}_{l+\hat{\mu}-\hat{\nu}\sigma})h_{l\sigma}\nonumber\\
&-&h^{\dagger}_{l\sigma}(h_{l+\hat{\mu}+\hat{\nu}\sigma} +
h_{l+\hat{\mu}-\hat{\nu}\sigma})]. \label{tcurpara}
\end{eqnarray}
For the convenience in the following discussions, the paramagnetic
component of the electron current operator can be rewritten in the
Nambu representation in terms of the charge carrier Nambu operators
$\Psi^\dagger_\vv{k}=\left ( h^{\dagger}_{\vv{k}\uparrow}, h_{-{\bf
k}\downarrow}\right)$ and $\Psi_{\vv{k}+\vv{q}} = \left(
h_{\vv{k}+\vv{q}\uparrow}, h^{\dagger}_{-\vv{k}-\vv{q}\downarrow}
\right)^T $. Moreover, since the density operator is summed over the
position of all particles, then its Fourier transform can be
obtained as $\rho(\vv{q})=({e}/2) \sum_{\vv{k}\sigma}
h^{\dagger}_{\vv{k}\sigma} h_{\vv{k}+\vv{q}\sigma} =({e}/2)
\sum_{\vv{k}} \Psi^{\dagger}_{\vv{k}}\tau_{3} \Psi_{\vv{k}+\vv{q}}$.
In this Nambu representation, the paramagnetic four-current operator
can be represented as,
\begin{eqnarray}\label{curnambu}
j_{\mu}^{(p)}(\vv{q})={1\over N}\sum\limits_{\vv{k}\sigma}
\Psi^{\dagger}_{\vv{k}}{\mathbf{\gamma}}_{\mu}(\vv{k},\vv{k}+\vv{q})
\Psi_{\vv{k}+\vv{q}}.
\end{eqnarray}
where the bare current vertex,
\begin{widetext}
\begin{eqnarray}
{\mathbf{\gamma}}_\mu(\vv{k}+\vv{q},\vv{k})= \left
\{\begin{array}{ll} -{2e\over\hbar}\, e^{{1\over 2}iq_{\mu}}
\{\sin(k_{\mu}+{1\over 2}q_{\mu})[\chi_{1}t-2\chi_{2}t'
\sum\limits_{\nu\neq\mu}\cos({1\over 2}q_{\nu})\cos(
k_{\nu}+{1\over 2}q_{\nu})]\\
-i(2\chi_{2}t')\cos(k_{\mu}+{1\over 2}q_{\mu})
\sum\limits_{\nu\neq\mu}\sin q_{\nu}\sin(k_{\nu}+{1\over 2}q_{\nu})
\}\tau_0
  & {\rm{for}}\ \mu\neq 0,~~~~\\
{e\over 2}\tau_3  & {\rm{for}}\ \mu=0.\\
\end{array}\right.\label{barevertex}
\end{eqnarray}
\end{widetext}

As in the previous discussions \cite{krzyzosiak10}, we are
calculating the polarization bubble with the paramagnetic current
operator (\ref{curnambu}), i.e., bare current vertices
(\ref{barevertex}), but Green function (\ref{holegreenfunction}). As
a consequence, we do not take into account longitudinal excitations
properly \cite{schrieffer83,misawa94}, the obtained results are
valid only in the gauge, where the vector potential is purely
transverse, e.g. in the Coulomb gauge. In this case, the correlation
function (\ref{corP}) can be obtained in the Nambu representation
as,
\begin{widetext}
\begin{eqnarray}
P_{\mu\nu}(\vv{q},i\omega_n) &=& {1\over N}\sum\limits_{\vv{k}}
{\mathbf{\gamma}}_\mu(\vv{k}+\vv{q},\vv{k})
{\mathbf{\gamma}}^{*}_\nu(\vv{k}+\vv{q},\vv{k}){1\over\beta}\sum
\limits_{i\nu_m}{\rm{Tr}}\left[{\mathbb{G}}(\vv{k+q},i\omega_n+i\nu_m)
{\mathbb{G}}(\vv{k},i\nu_m)\right].~~~~\label{barepolmats}
\end{eqnarray}
Substituting the Green's function (\ref{holegreenfunction}) into Eq.
(\ref{barepolmats}), we then obtain the paramagnetic part of the
response kernel in the static limit ($\omega\sim 0$) as,
\begin{eqnarray}
K_{\mu\nu}^{(\rm{p})}(\vv{q},0) &=&{1\over N}\sum\limits_{\vv{k}}
{\mathbf{\gamma}}_\mu(\vv{k}+\vv{q},\vv{k})
{\mathbf{\gamma}}^{*}_\nu(\vv{k}+\vv{q},\vv{k})[L_{1}(\vv{k},\vv{q})+
L_{2}(\vv{k},\vv{q})]=K_{\mu\mu}^{(\rm{p})}(\vv{q},0)\delta_{\mu\nu},
~~~~\label{parakernel}
\end{eqnarray}
with the functions $L_{1}(\vv{k},\vv{q},\omega)$ and
$L_{2}(\vv{k},\vv{q},\omega)$ are given by,
\begin{subequations}\label{lfunctions}
\begin{eqnarray}
L_{1}(\vv{k},\vv{q})&=&Z^{2}_{\rm{hF}}\left(1+ {\xikq\xik +
\Dkq\Dk\over \Ek\Ekq}\right){\nF(\Ek)-\nF(\Ekq)\over\Ek-\Ekq},\\
L_{2}(\vv{k},\vv{q})&=&Z^{2}_{\rm{hF}}\left(1- {\xikq\xik +
\Dkq\Dk\over \Ek\Ekq}\right){\nF(\Ek)+\nF(\Ekq)-1\over\Ek+\Ekq},
~~~~~~~
\end{eqnarray}
\end{subequations}
\end{widetext}
respectively. In this case, the kernel of the response function in
Eq. (\ref{kernel}) is now obtained from Eqs. (\ref{diakernel}) and
(\ref{parakernel}) as,
\begin{eqnarray}
K_{\mu\nu}(\vv{q},0)=\left [{1\over\lambda^{2}_{L}}+
K_{\mu\mu}^{(\rm{p})}(\vv{q},0)\right ]\delta_{\mu\nu}.
\label{kernel1}
\end{eqnarray}
It should be emphasized that in the present CSS fermion-spin theory
\cite{feng0304,feng07}, these charge carrier $h^{\dagger}_{l\sigma}$
and spin ${\bf S}_{l}$ are gauge invariant, and in this sense they
are real and can be interpreted as physical excitations
\cite{laughlin97}. Furthermore, as shown in Eq. (\ref{poloper}) and
Eq. (\ref{tcurpara}), the electron polarization operator and the
related electron current operator are identified with the
corresponding charge carrier polarization operator and charge
carrier current operator, since the electron single occupancy local
constraint is satisfied in the CSS fermion-spin approach.

\subsection{Doping dependence of the Meissner effect in the long
wavelength limit}

\begin{figure}[h!]
\includegraphics[scale=0.32]{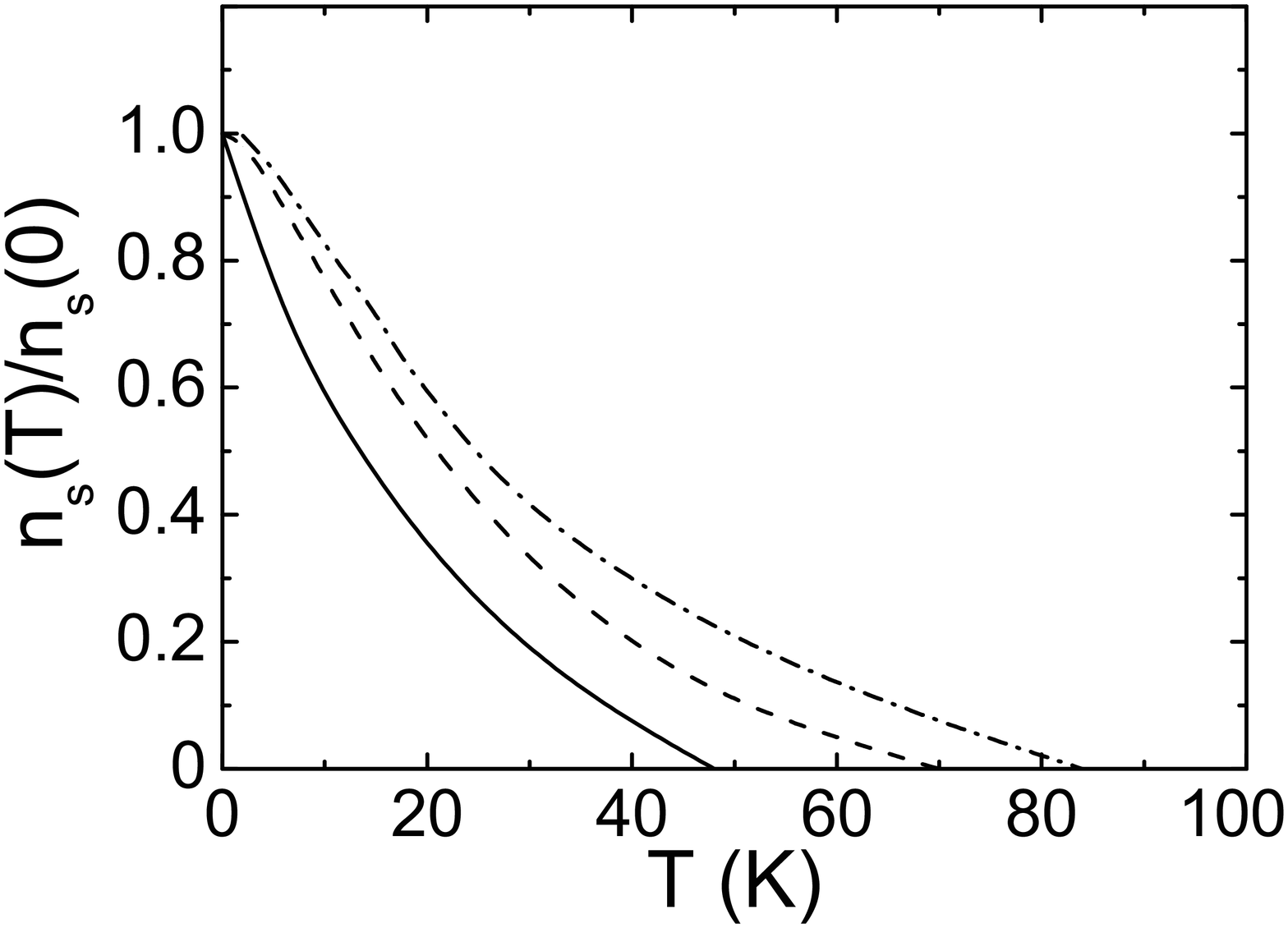}
\caption{The effective superfluid density as a function of
temperature $T$ for the doping concentration $\delta=0.09$ (solid
line), $\delta=0.12$ (dashed line), and $\delta=0.15$ (dash-dotted
line) with parameters $t/J=2.5$, $t'/t=0.3$, and $J=1000$K.
\label{effectivsuperfluid}}
\end{figure}

With the help of the response kernel function (\ref{kernel1}), we
now discuss the doping and temperature dependence of the Meissner
effect in cuprate superconductors. In particular, in the long
wavelength limit, i.e., $|\vv{q}|\to 0$, the function
$L_{2}(\vv{k},\vv{q}\to 0)$ vanishes, then the paramagnetic part of
the response kernel can be obtained explicitly as,
\begin{widetext}
\begin{eqnarray}
K_{yy}^{(\rm{p})}(\vv{q}\to 0,0) &=&2Z^{2}_{\rm{hF}} {4e^{2}\over
\hbar^{2}}{1\over N}\sum\limits_{\vv{k}}\sin^{2}k_{y}[\chi_{1}t-
2\chi_{2}t'\cos k_{x}]^{2}\lim\limits_{\vv{q}\to 0}
{\nF(\Ek)-\nF(\Ekq)\over\Ek-\Ekq}.~~~~~~ \label{kernel5}
\end{eqnarray}
\end{widetext}
At zero-temperature $T=0$, $K_{yy}^{(\rm{p})}(\vv{q}\to 0,0)|_{T=0}
=0$ (see Appendix \ref{limit1}). In this case, the long wavelength
electromagnetic response is determined by the diamagnetic part of
the kernel only. On the other hand, at the SC transition temperature
$T=T_{c}$, the effective charge carrier gap parameter
$\bar{\Delta}_{\rm h}|_{T=T_{c}}=0$. In this case, the paramagnetic
part of the response kernel in the long wavelength limit can be
evaluated as (see Appendix \ref{limit1}),
\begin{widetext}
\begin{eqnarray}
K_{yy}^{(\rm{p})}(\vv{q}\to 0,0)|_{T=T_{c}} &=&2Z^{2}_{\rm{hF}}
{4e^{2}\over \hbar^{2}}{1\over N}\sum\limits_{\vv{k}} \sin^{2}k_{y}
[\chi_{1}t- 2\chi_{2}t'\cos k_{x}]^{2}\lim\limits_{\vv{q}\to 0}
{\nF(\xik)-\nF(\xikq)\over\xik-\xikq}= -{1\over \lambda^{2}_{L}},
\end{eqnarray}
\end{widetext}
which exactly cancels the diamagnetic part of the response kernel
(\ref{diakernel}), and then the Meissner effect in cuprate
superconductors is obtained for all $T\leq T_{c}$. To show this
point clearly, we define the effective superfluid density $n_{s}(T)$
at temperature $T$ in terms of the paramagnetic part of the response
kernel as,
\begin{eqnarray}
K_{\mu\nu}^{(\rm{p})}(\vv{q}\to 0,0)=-{1\over\lambda^{2}_{L}}\left
[1- {n_{s}(T)\over n_{s}(0)}\right ]\delta_{\mu\nu},
\end{eqnarray}
and then the kernel of the response function in Eq. (\ref{kernel1})
can be rewritten as,
\begin{eqnarray}
K_{\mu\nu}(\vv{q}\to 0,0)={1\over\lambda^{2}_{L}}{n_{s}(T)\over
n_{s}(0)}\delta_{\mu\nu}, \label{kernel2}
\end{eqnarray}
where the ratio $n_{s}(T)/n_{s}(0)$ of the effective superfluid
densities at temperature $T$ and zero-temperature is given by,
%\begin{widetext}
\begin{eqnarray}\label{ratio2}
{n_{s}(T)\over n_{s}(0)}&=&1-2\lambda^{2}_{L}Z^{2}_{\rm{hF}}{4e^{2}
\over \hbar^{2}}{1\over N}\sum\limits_{\vv{k}}\sin^{2}k_{y}
[\chi_{1}t-2\chi_{2}t'\cos k_{x}]^{2}\nonumber\\
&\times&{\beta e^{\beta\Ek}\over (e^{\beta\Ek}+1)^{2}}.
\end{eqnarray}
%\end{widetext}
In cuprate superconductors, although the values of $J$, $t$, and
$t'$ are believed to vary somewhat from compound to compound
\cite{damascelli03}, however, as a qualitative discussion, the
commonly used parameters in this paper are chosen as $t/J=2.5$,
$t'/t=0.3$, and $J=1000$K. In this case, we plot the effective
superfluid density $n_{s}(T)/n_{s}(0)$ as a function of temperature
$T$ for the doping concentration $\delta=0.09$ (solid line),
$\delta=0.12$ (dashed line), and $\delta=0.15$ (dash-dotted line) in
Fig. \ref{effectivsuperfluid}. Obviously, it is shown that the
effective superfluid density diminishes with increasing
temperatures, and disappears at the SC transition temperature
$T_{c}$, then all the charge carriers are in the normal fluid for
the temperature $T\geq T_{c}$. In summary, we have found the
following results within the kinetic energy driven SC mechanism: (1)
the doping dependence of the Meissner effect in cuprate
superconductors is obtained for all temperature $T\leq T_{c}$
throughout the SC dome; (2) the electromagnetic response kernel goes
to the London form in the long wavelength limit [see Eq.
(\ref{kernel2})]; (3) although the electromagnetic response kernel
is not manifestly gauge invariant within the present bare current
vertex (\ref{barevertex}), it has been shown that the gauge
invariance is kept within the dressed current vertex
\cite{krzyzosiak10}.

\section{Quantitative characteristics of the Meissner effect}
\label{meissner}

\begin{figure}[h!]
\includegraphics[scale=0.31]{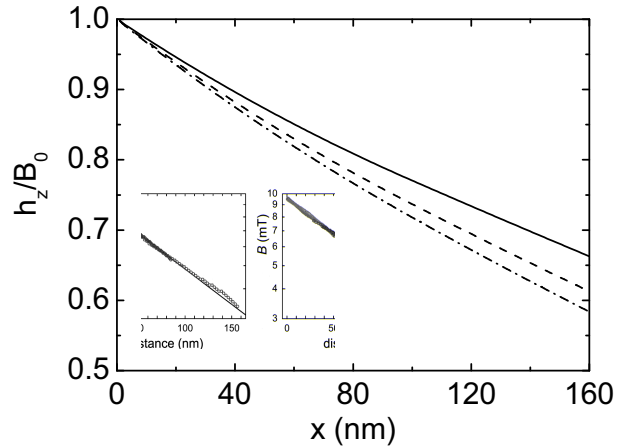}
\caption{The local magnetic field profile as a function of the
distance from the surface at temperature $T=2$K for the doping
concentration $\delta=0.09$ (solid line), $\delta=0.12$ (dashed
line), and $\delta=0.15$ (dash-dotted line) with parameters
$t/J=2.5$, $t'/t=0.3$, and $J=1000$K. Inset: the corresponding
experimental result for YBa$_2$Cu$_3$O$_{7-y}$ taken from Ref.
\onlinecite{suter04}. \label{profilefig}}
\end{figure}

In this section, we discuss the basic behavior of cuprate
superconductors in the presence of a weak electromagnetic field. As
seen from Eq. (\ref{linres}), once the response kernel $K_{\mu\nu}$
is known, the effect of a weak electromagnetic field can be
quantitatively characterized by experimentally measurable quantities
such as the local magnetic field profile, the magnetic field
penetration depth, and the superfluid density. However, the result
we have obtained the response kernel in Eq. (\ref{kernel1}) [then
the effective superfluid density (\ref{ratio2})] can not be used for
a direct comparison with the corresponding experimental data of
cuprate superconductors because the kernel function derived within
the linear response theory describes the response of an
\emph{infinite} system, whereas in the problem of the penetration of
the field and the system has a surface, i.e., it occupies a
half-space $x>0$. In such problems, it is necessary to impose
boundary conditions for charge carriers. This can be done within the
simplest specular reflection model \cite{landau80,tinkham96} with a
two-dimensional geometry of the SC plane. In this case, one may use
the response kernel $K_{\mu\nu}$ for an infinite space, however, it
is necessary to extend the vector potential $\vv{A}({\bf r})$ in an
even manner through the boundary. If the externally applied magnetic
field is perpendicular to the ab plane, we may choose $A_{y}(x)$
along the $y$ axis. Following the Maxwell equation,
\begin{eqnarray}
{\rm rot}\,{\bf B}={\rm rot}\,{\rm rot}\,{\bf A}={\rm grad}\,{\rm
div}\,\vv{A}-\nabla^2 \,\vv{A}=\mu_{0}{\bf J},
\end{eqnarray}
it is shown clearly that the extension of the vector potential in an
even manner through the boundary implies a kink in the $A_{y}(x)$
curve. In other words, if the externally applied magnetic field
${\bf B}$ is given at the system surface, i.e., $({\rm d}A_{y}(x)/
{\rm d}x)|_{x=+0}=B$, while $({\rm d}A_{y}(x)/{\rm d}x)
|_{x=-0}=-B$, this leads to a fact \cite{landau80} that the second
derivative $({\rm d}^{2}A_{y}(x)/ {\rm d}^{2}x)$ acquires a
correction $2B\delta(x)$, i.e.,
\begin{eqnarray}\label{correction}
{{\rm d}^{2}A_{y}(x)\over {\rm d}^{2}x}=2B\delta(x)-\mu_{0}J_{y},
\end{eqnarray}
where the transverse gauge ${\rm div}\,\vv{A}=0$ has been adopted.
In the momentum space, this equation (\ref{correction}) can be
expressed as $q_{x}^{2}A_{y}(\vv{q})=\mu_{0}J_{y}(\vv{q})-2B$.
Substituting this Fourier transform of Eq. (\ref{correction}) into
Eq. (\ref{linres}), and solving for the vector potential we obtain,
\begin{equation}\label{aspec}
A_y(\vv{q})=-2B{\delta(q_y)\delta(q_z)\over\mu_{0}K_{yy}(\vv{q})
+q_{x}^{2}}.
\end{equation}
Since the vector potential has only the $y$ component, the non-zero
component of the local magnetic field $\vv{h} = \rm{rot}\,\vv{A}$ is
that along the $z$ axis and $h_{z}(\vv{q})= iq_{x}A_{y}(\vv{q})$.
With the help of Eq. (\ref{aspec}), we can obtain explicitly the
local magnetic field profile as,
\begin{equation}\label{profile}
h_{z}(x)={B\over\pi}\int\limits_{-\infty}^\infty {\rm{d}}q_{x}\,
{q_{x}\sin(q_{x}x)\over\mu_{0}K_{yy}(q_{x},0,0) + q_{x}^{2}},
\end{equation}
which therefore reflects the measurably electromagnetic response in
cuprate superconductors. For the convenience in the following
discussions, we introduce a characteristic length scale
$a_{0}=\sqrt{\hbar^{2}a/\mu_{0}e^{2}J}$. Using the lattice parameter
$a\approx 0.383$nm for the cuprate superconductor
YBa$_2$Cu$_3$O$_{7-y}$, this characteristic length is obtain as
$a_{0}\approx 97.8$nm. In Fig. \ref{profilefig}, we plot the local
magnetic field profile (\ref{profile}) as a function of the distance
from the surface at temperature $T=2$K for the doping concentration
$\delta=0.09$ (solid line), $\delta=0.12$ (dashed line), and
$\delta=0.15$ (dash-dotted line) in comparison with the
corresponding experimental result \cite{suter04} of the local
magnetic field profiles for the high quality YBa$_2$Cu$_3$O$_{7-y}$
(inset). If an external field $B=8.82$mT is applied to the system
just as it has been done in the experimental measurement
\cite{suter04}, then the experimental result \cite{suter04} for
YBa$_2$Cu$_3$O$_{7-y}$ is well reproduced. In particular, our
theoretical results perfectly follow an exponential field decay as
expected for the local electrodynamic response. This is different
from the conventional superconductors, where the local magnetic
field profile in the Meissner state shows a clear deviation from the
exponential field decay \cite{suter04}. The exponential character of
the local magnetic field profile has been observed experimentally on
different families of cuprate superconductors
\cite{khasanov04,suter04}, in support of a local (London-type)
nature of the electrodynamics \cite{schrieffer83}.

\subsection{Doping and temperature dependence of the magnetic field
penetration depth}\label{penetrationdepth}

\begin{figure}[h!]
\includegraphics[scale=0.31]{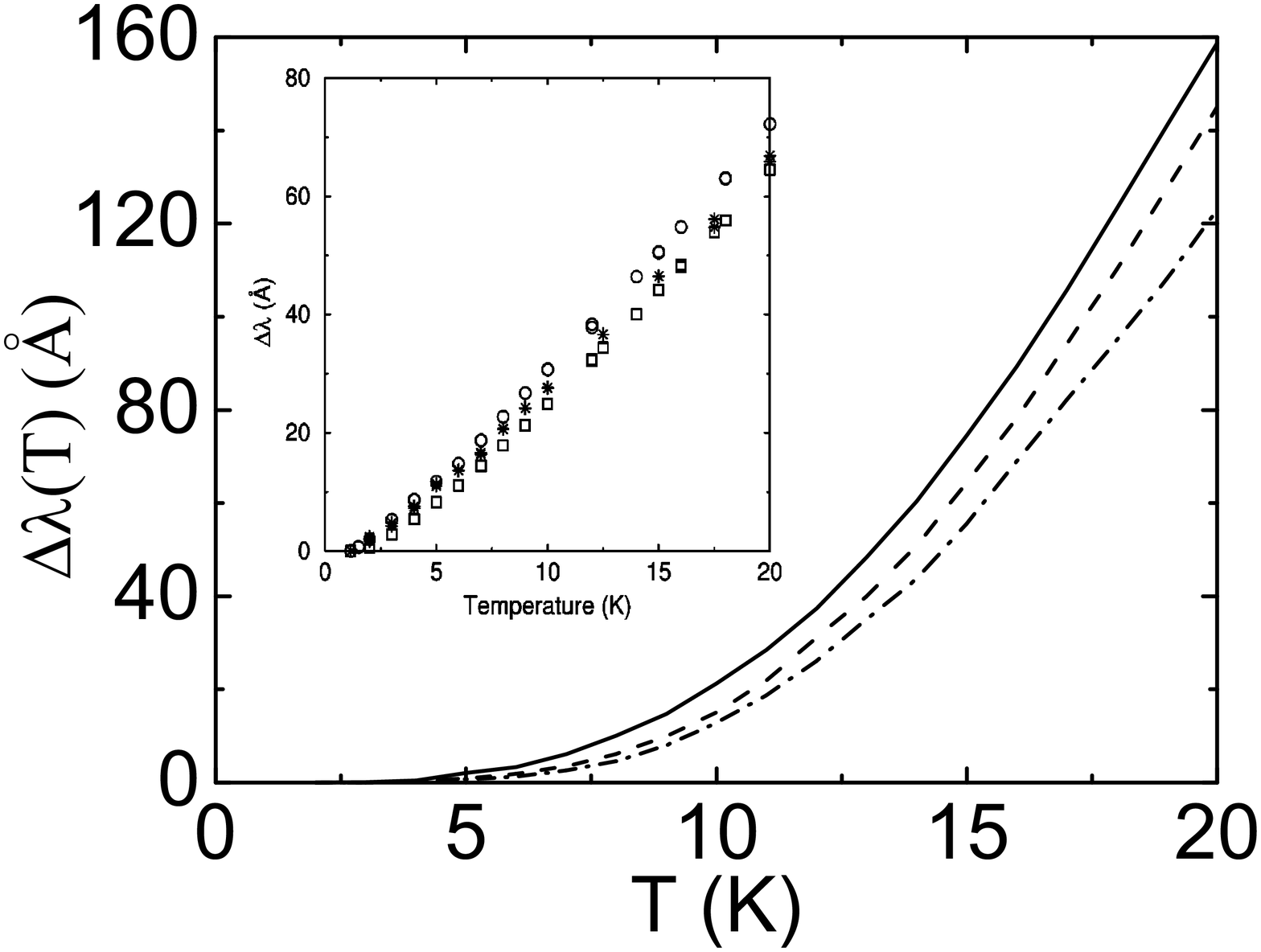}
\caption{Temperature dependence of the magnetic field penetration
depth $\Delta\lambda(T)$ for the doping concentration $\delta=0.14$
(solid line), $\delta=0.15$ (dashed line), and $\delta=0.18$
(dash-dotted line) with parameters $t/J=2.5$, $t'/t=0.3$, and
$J=1000$K. Inset: the corresponding experimental data for
YBa$_2$Cu$_3$O$_{7-y}$ taken from Ref. \onlinecite{kamal98}.
\label{lambdafig}}
\end{figure}

In the following discussions, we do not analyze the behavior of the
filed in the general case, and only discuss the doping and
temperature dependence of the magnetic field in-plane penetration
depth $\lambda(T)$ and the related in-plane superfluid density
$\rho_{\rm s}(T)$. The magnetic field in-plane penetration depth is
defined in terms of the local magnetic field profile (\ref{profile})
as,
\begin{eqnarray}\label{lambda}
\lambda(T)={1\over B}\int\limits_{0}^{\infty}h_{z}(x)\,{\rm{d}}x=
{2\over\pi}\int\limits_{0}^{\infty}{\rm{d}q_x\over\mu_{0}
K_{yy}(q_{x},0,0)+q_{x}^{2}}.~~~~~
\end{eqnarray}
At zero-temperature, the calculated magnetic field penetration
depths are $\lambda(0)\approx 239.17$nm, $\lambda(0)\approx
234.76$nm, and $\lambda(0)\approx 224.44$nm for the doping
concentrations $\delta=0.14$, $\delta=0.15$, and $\delta=0.18$,
respectively, which are consistent with the values of the magnetic
field penetration depth $\lambda\approx 156$nm $\sim 400$nm observed
for different families of cuprate superconductors in different
doping concentrations
\cite{bernhard01,khasanov04,niedermayer93,broun07,uemura93}. On the
other hand, at the SC transition temperature $T=T_{c}$, the kernel
of the response function $K_{\mu\nu}(\vv{q}\to 0,0)|_{T=T{c}}=0$. In
this case, we obtain the magnetic field penetration depth from Eq.
(\ref{lambda}) as $\lambda(T_{c})=\infty$, which reflects that in
the normal state, the external magnetic field can penetrate through
the main body of the system, therefore there is no the Meissner
effect in the normal state. For a better understanding of the
unusual behavior of the temperature dependence of the magnetic field
penetration depth $\lambda(T)$, $\Delta\lambda(T)=\lambda(T)-
\lambda(0)$ as a function of temperature $T$ for the doping
concentration $\delta=0.14$ (solid line), $\delta=0.15$ (dashed
line), and $\delta=0.18$ (dash-dotted line) is plotted in Fig.
\ref{lambdafig} in comparison with the corresponding experimental
results \cite{kamal98} of YBa$_2$Cu$_3$O$_{7-y}$ (inset). Our result
shows clearly that in low temperature, the magnetic field
penetration depth $\Delta\lambda(T)$ exhibits a linear temperature
dependence, however, it crosses over to a nonlinear behavior in the
extremely low temperatures, in good agreement with experimental
observation in nominally clean crystals of cuprate superconductors
\cite{hardy93,kamal98,jackson00,pereg07,khasanov04,suter04}. In
comparison with our previous discussions \cite{krzyzosiak10}, our
present results also show that the good agreement can be reached by
introducing the second-nearest neighbors hopping $t'$ in the nearest
neighbors hopping $t$-$J$ model. It should be emphasized that the
present result for cuprate superconductors is much different from
that in the conventional superconductors, where the characteristic
feature is the existence of the isotropic energy gap $\Delta_{s}$,
and then $\Delta\lambda(T)$ exhibits an exponential behavior as
$\Delta\lambda(T)\propto {\rm exp}(-\Delta_{s}/T)$.

\subsection{Doping and temperature dependence of the in-plane
superfluid density}

\begin{figure}[h!]
\includegraphics[scale=0.32]{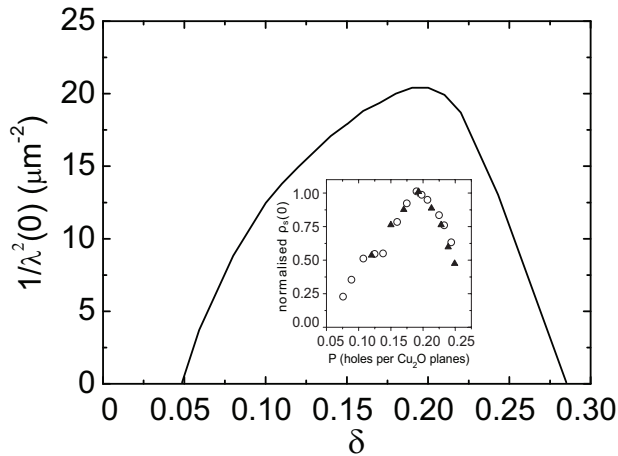}
\caption{Doping dependence of the zero-temperature superfluid
density with parameters $t/J=2.5$, $t'/t=0.3$, and $J=1000$K. Inset:
the corresponding experimental result for
Y$_{0.8}$Ca$_{0.2}$Ba$_{2}$(Cu$_{1-z}$Zn$_{z}$)$_{3}$O$_{7-\delta}$
(open circles) and
Tl$_{1-y}$Pb$_{y}$Sr$_{2}$Ca$_{1-x}$Y$_x$Cu$_2$O$_{7}$ (solid
triangles) taken from Ref. \onlinecite{bernhard01}. \label{rhofig}}
\end{figure}

Now we turn to discuss the doping and temperature dependence of the
in-plane superfluid density $\rho_{\rm s}(T)$, which is a measure of
the phase stiffness, and can be obtained in terms of the magnetic
field in-plane penetration depth $\lambda(T)$ as,
\begin{eqnarray}
\rho_{\rm s}(T)\equiv {1\over \lambda^{2}(T)}. \label{rhodensity}
\end{eqnarray}
In this case, we have performed firstly a calculation for the doping
dependence of the zero-temperature superfluid density $\rho_{\rm s}
(0)$ for all levels of doping, and the result is plotted in Fig.
\ref{rhofig} in comparison with the corresponding experimental data
\cite{bernhard01} for
Y$_{0.8}$Ca$_{0.2}$Ba$_{2}$(Cu$_{1-z}$Zn$_{z}$)$_{3}$O$_{7-\delta}$
and Tl$_{1-y}$Pb$_{y}$Sr$_{2}$Ca$_{1-x}$Y$_x$Cu$_2$O$_{7}$ (inset).
It is shown clearly that the zero-temperature superfluid density
increases with increasing doping in the lower doped regime, and
reaches a maximum (a peak) around the critical doping $\delta\approx
0.195$, then decreases in the higher doped regime, in good agreement
with the experimental results of cuprate superconductors
\cite{niedermayer93,bernhard01,broun07,khasanov10}. Since the
superfluid density $\rho_{\rm s}(T)$ [then the magnetic field
penetration depth $\lambda(T)$] is related to the current-current
correlation function, the charge carrier pair gap parameter is
relevant as shown in Eqs. (\ref{parakernel}), (\ref{lfunctions}),
and (\ref{kernel1}), i.e., the variation of the superfluid density
with doping and temperature is coupled to the doping and temperature
dependence of the charge carrier pair gap parameter
$\bar{\Delta}_{hZ}$ in cuprate superconductors. In this case, our
present domelike shape of the doping dependent superfluid density
also is a natural consequence of the domelike shape of the doping
dependence of the SC transition temperature (then the charge carrier
pair gap parameter) in the framework of the kinetic energy driven SC
mechanism \cite{feng0306}, where the the maximal SC transition
temperature (then the charge carrier pair gap parameter) occurs in
the optimal doping, and then decreases in both underdoped and
overdoped regimes. However, the calculated SC transition temperature
$T_{c}$ [then the zero-temperature gap parameter $\bar{\Delta}_{\rm
h}(0)$] exhibits the maximal value at the optimal doping
$\delta_{\rm optimal}\approx 0.15$ \cite{guo07}, therefore there is
a difference between the optimal doping $\delta_{\rm optimal}\approx
0.15$ [the maximal $\bar{\Delta}_{\rm h}(0)$ value] and the critical
doping $\delta_{\rm critical}\approx 0.195$ [the highest $\rho_{\rm
s}(0)$ value]. This difference can be understood within the present
theoretical framework. In the domelike shape of the doping
dependence of the gap parameter $\bar{\Delta}_{\rm h}(0)$, the gap
parameter $\bar{\Delta}_{\rm h}(0)$ reaches its maximal value at the
optimal doping $\delta_{\rm optimal}\approx 0.15$, where the
doping-derivative of $\bar{\Delta}_{\rm h}(0)$ can be obtained as
$({\rm d} \bar{\Delta}_{\rm h}(0)/{\rm d}\delta)|_{\delta=
\delta_{\rm optimal}}=0$. On the other hand, at the critical doping
$\delta_{\rm critical}\approx 0.195$, the peak of the superfluid
density $\rho_{\rm s}(0)$ appears, where the doping-derivative of
$\rho_{\rm s}(0)$ can be obtained as $({\rm d}\rho_{\rm s}(0)/{\rm
d} \delta)|_{\delta=\delta_{\rm critical}}=0$. According to the
definition of the superfluid density $\rho_{\rm s}(T)$ in Eq.
(\ref{rhodensity}), $({\rm d}\rho_{\rm s}(0)/{\rm d}\delta)
|_{\delta=\delta_{\rm critical}}=0$ is equivalent to $({\rm
d}\lambda(0)/{\rm d}\delta)|_{\delta=\delta_{\rm critical}}=0$. In
this case, $({\rm d}\lambda(0)/{\rm d}\delta)|_{\delta= \delta_{\rm
critical}}=0$ can be expressed from Eq. (\ref{lambda}) as,
\begin{widetext}
\begin{eqnarray}\label{dflambda}
\left [{{\rm d}\lambda(0)\over {\rm d}\delta}\right ]_{\delta=
\delta_{\rm critical}}=- {2\mu_{0}\over\pi}\int\limits_{0}^{\infty}
\rm{d}q_{x}\left [{1\over [\mu_{0} K_{yy}(q_{x},0,0)+q_{x}^{2}]^{2}}
{{\rm d}K_{yy}(q_{x},0,0)\over {\rm d}\delta}\right
]_{\delta=\delta_{\rm critical}}=0,
\end{eqnarray}
\end{widetext}
then with the help of Eqs. (\ref{diakernel}), (\ref{parakernel}),
and (\ref{lfunctions}), it is straightforward to find that when
$({\rm d}\rho_{\rm s}(0)/{\rm d}\delta)|_{\delta=\delta_{\rm
critical}}=0$, $({\rm d}\bar{\Delta}_{\rm h} (0)/{\rm d}
\delta)|_{\delta=\delta_{\rm critical}}\neq 0$, since the
diamagnetic part of the response kernel in Eq. (\ref{diakernel}),
the spin correlation functions $\chi_{1}=\langle S_{i}^{+}
S_{i+\hat{\eta}}^{-}\rangle$ and $\chi_{2}=\langle S_{i}^{+}
S_{i+\hat{\eta}'}^{-}\rangle$, and the charge carrier particle-hole
parameters $\phi_{1}=\langle h^{\dagger}_{i\sigma}
h_{i+\hat{\eta}\sigma}\rangle$ and $\phi_{2}= \langle
h^{\dagger}_{i\sigma}h_{i+\hat{\eta}'\sigma} \rangle$ are also
doping dependent. This leads to the difference between the optimal
doping $\delta_{\rm optimal}\approx 0.15$ for the zero-temperature
gap parameter (then the SC transition temperature) and the critical
doping $\delta_{\rm critical}\approx 0.195$ for the zero-temperature
superfluid density. In particular, it is found that $({\rm d}
\bar{\Delta}_{\rm h}(0)/{\rm d}\delta)|_{\delta=\delta_{\rm
critical}}< 0$, indicating that the critical doping locates at the
slightly overdoped regime.  Furthermore, it should be emphasized
that the early experimental data observed from cuprate
superconductors show that the superfluid density $\rho_{\rm s}(0)$
in the underdoped regime vanishes more or less linearly with
decreasing doping concentration \cite{uemura8991}. Later, a clear
deviation from this linear relation between the superfluid density
$\rho_{\rm s}(0)$ and doping concentration has been observed in the
underdoped regime \cite{pereg07,niedermayer93,bernhard01,broun07}.
In particular, the recent experimental measurement \cite{khasanov10}
on the cuprate superconductor YBa$_2$Cu$_3$O$_{7-y}$ indicate that
the superfluid density $\rho_{\rm s}(0)$ is, in actual fact,
linearly proportional to the doping concentration in the low doping
range ($\delta\approx 0.054\sim 0.061$). Our present result in the
low doping range also is well consistent with this recent
experimental observation \cite{khasanov10}.

\begin{figure}[h!]
\includegraphics[scale=0.33]{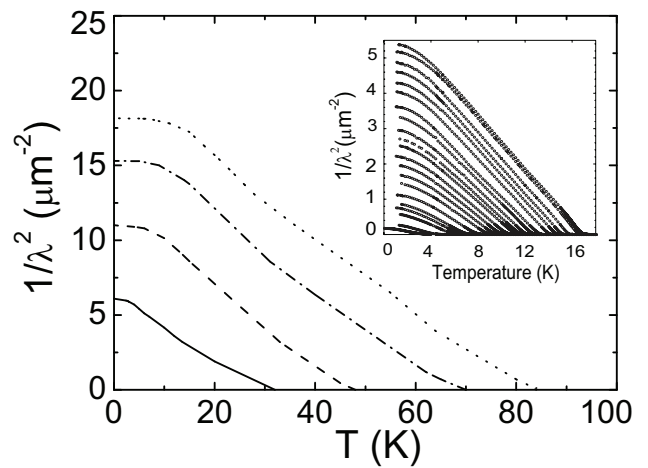}
\caption{Temperature dependence of the superfluid density for the
doping concentration $\delta=0.06$ (solid line), $\delta=0.09$
(dashed line), $\delta=0.12$ (dash-dotted line), and $\delta=0.15$
(dotted line) with parameters $t/J=2.5$, $t'/t=0.3$, and $J=1000$K.
Inset: the corresponding experimental result for
YBa$_2$Cu$_3$O$_{7-y}$ taken from Ref. \onlinecite{khasanov10}.
\label{rhofigt}}
\end{figure}

The doping dependence of the superfluid density shown in Fig.
\ref{rhofig} also is strongly temperature dependent. In particular,
when the temperature $T=T_{c}$, the kernel of the response function
$K_{\mu\nu}(\vv{q}\to 0,0)|_{T=T{c}} =0$ and the magnetic field
penetration depth $\lambda(T_{c})=\infty$ as mentioned in Subsection
\ref{penetrationdepth}, this leads to the superfluid density
$\rho_{\rm s}(T_{c})=0$, which is consistent with the result of the
effective superfluid density obtained from Eq. (\ref{ratio2}). To
show the superfluid density clearly for all the temperature $T\leq
T_{c}$, we plot the superfluid density $\rho_{\rm s}(T)$ as a
function of temperature for the doping concentration $\delta=0.06$
(solid line), $\delta=0.09$ (dashed line), $\delta=0.12$
(dash-dotted line), and $\delta=0.15$ (dotted line) in Fig.
\ref{rhofigt} in comparison with the corresponding experimental
result \cite{khasanov10} of YBa$_2$Cu$_3$O$_{7-y}$ (inset). Our
result indicates that the superfluid density $\rho_{\rm s}(T)$
decreases with increasing temperature, and vanishes at the SC
transition temperature $T_{c}$. Moreover, the most striking feature
of the present results is the wide range of linear temperature
dependence at low temperature, extending from close to the SC
transition temperature $T_{c}$ to down to the temperatures $T\approx
4$K$\sim 8$K for different doping concentrations. However, in
correspondence with the nonlinear temperature dependence of the
magnetic field penetration depth  at the extremely low temperatures
shown in Fig. \ref{lambdafig}, the superfluid density $\rho_{\rm
s}(T)$ crosses over to a nonlinear temperature behavior at the
extremely low temperatures (below $T\approx 4$K$\sim 8$K for
different doping concentrations). Our these results are also well
consistent with the corresponding experimental result
\cite{niedermayer93,bernhard01,broun07,khasanov10} for cuprate
superconductors. The good agreement between the present theoretical
results and experimental data also shows that the d-wave
contribution to the superfluid density $\rho_{\rm s}(T)$ in cuprate
superconductors is predominant.

The essential physics of the nonlinearity in the temperature
dependence of the penetration depth (then the superfluid density) in
cuprate superconductors at the extremely low temperatures, as shown
in Fig. \ref{lambdafig} (then Fig. \ref{rhofigt}), in the present
$t$-$t'$-$J$ model is the same as that in the $t$-$J$
\cite{krzyzosiak10}, and can be attributed to the nonlocal effects
induced by the gap nodes on the Fermi surface in a pure d-wave
pairing state \cite{yip92,kosztin97,franz97,li00,sheehy04}. A weak
external magnetic field acts on the SC state of cuprate
superconductors as a perturbation. Within the linear response
theory, one can find that the nonlocal relation between the
supercurrent and the vector potential (\ref{linres}) in the
coordinate space holds due to the finite size of Cooper pairs. In
particular, in the present kinetic energy driven d-wave SC mechanism
\cite{feng0306}, the size of charge carrier pairs in the clean limit
is of the order of the coherence length $\zeta(\vv{k})=\hbar v_{\rm
F}/\pi \Delta_{\rm h}(\vv{k})$, where $v_{\rm{F}}=\hbar^{-1}
\partial\xi_{\bf k}/\partial {\bf k}|_{k_{F}}$ is the charge carrier
velocity at the Fermi surface, which shows that the size of charge
carrier pairs is momentum dependent. In general, although the
external magnetic field decays exponentially on the scale of the
magnetic field penetration length $\lambda(T)$, any nonlocal
contributions to measurable quantities are of the order of
$\kappa^{-2}$, where the Ginzburg--Landau parameter $\kappa$ is the
ratio of the magnetic field penetration depth $\lambda$ and the
coherence length $\zeta$. However, for cuprate superconductors,
because the pairing is d-wave, the charge carrier gap vanishes on
the gap nodes on the Fermi surface, so that the quasiparticle
excitations are gapless and therefore affect particularly the
physical properties at the extremely low temperatures. This gapless
quasiparticle excitation leads a divergence of the coherence length
$\zeta(\vv{k})$ around the gap nodes on the Fermi surface, and then
the behavior of the temperature dependence of the magnetic field
penetration depth (then the superfluid density) depends sensitively
on the quasiparticle scattering. At the extremely low temperatures,
the quasiparticles selectively locate around the gap nodal region,
and then the major contribution to measurable quantities comes from
these quasiparticles. In this case, the Ginzburg--Landau ratio
$\kappa(\vv{k})$ around the gap nodal region is no longer large
enough for the system to belong to the class of type-II
superconductors, and the condition of the local limit is not
satisfied \cite{kosztin97}, which leads to the system in the extreme
nonlocal limit, and therefore the nonlinear behavior in the
temperature dependence of the magnetic field penetration depth (then
superfluid density) is observed experimentally
\cite{bonn96,khasanov04,suter04}. On the other hand, with increasing
temperature, the quasiparticles around the gap nodal region become
excited out of the condensate, and then the nonlocal effect fades
away, where the momentum dependent coherence length $\zeta(\vv{k})$
can be replaced approximately with the isotropic one $\zeta_0=\hbar
v_{\rm F}/ \pi \Delta_{\rm h}$. In this case, the calculated
Ginzburg--Landau parameters are $\kappa_0\approx \lambda(0)/\zeta_0
\approx 166.29$, $\kappa_0\approx 175.55$, and $\kappa_0\approx
156.14$ for the doping concentrations $\delta=0.14$, $\delta=0.15$,
and $\delta=0.18$, respectively, and therefore the condition for the
local limit is satisfied. In particular, these theoretical values of
the Ginzburg--Landau parameter in different doping concentrations
are very close to the range $\kappa_0 \approx 150\sim 400$ estimated
experimentally for different families of cuprate superconductors in
different doping concentrations
\cite{bernhard01,khasanov04,niedermayer93,broun07,uemura93}. As a
consequence, our present study shows that cuprate superconductors at
moderately low temperatures turn out to be type-II superconductors,
where nonlocal effects can be neglected, then the electrodynamics is
purely local and the magnetic field decays exponentially over a
length of the order of a few hundreds nm.

\section{Conclusions}
\label{conclusions}

Within the $t$-$t'$-$J$ model, we have discussed the doping and
temperature dependence of the Meissner effect in cuprate
superconductors based on the kinetic energy driven SC mechanism. Our
results show that in the linear response approach, the
electromagnetic response consists of two parts as in the
conventional superconductors, the diamagnetic current, which is the
acceleration in the magnetic field, and the paramagnetic current,
which is a perturbation response of the excited quasiparticle and
exactly cancels the diamagnetic term in the normal state, then the
Meissner effect is obtained for all the temperature $T\leq T_{c}$
throughout the SC dome. Within this framework, we have reproduced
well all the main features of the doping dependence of the local
magnetic field profile, the magnetic field penetration depth, and
the superfluid density in terms of the specular reflection model.
The local magnetic field profile follows an exponential law, while
the magnetic field penetration depth shows a crossover from the
linear temperature dependence at low temperatures to a nonlinear one
at the extremely low temperatures. Moreover, in analogy to the
domelike shape of the doping dependent SC transition temperature,
the superfluid density increases with increasing doping in the lower
doped regime, and reaches a maximum around the critical doping
$\delta\approx 0.195$, then decreases in the higher doped regime.
The good agreement between the present theoretical results in the
clean limit and experimental data for different families of cuprate
superconductors also provides an important confirmation of the
nature of the SC phase of cuprate superconductors as a d-wave
BCS-like SC state within the kinetic energy driven SC mechanism.

Finally, it should be emphasized that in the present study, the only
coupling of the electron charge to the weak external magnetic field
is considered in terms of the vector potential $\vv{A}$, while the
coupling of the electron magnetic momentum with the weak external
magnetic field in terms of the Zeeman mechanism has been dropped. In
this case, the above obtained results are only valid in the weak
external magnetic field limit. However, the depairing due to the
Pauli spin polarization is very important in the presence of a
moderate or strong external magnetic field, since cuprate
superconductors are doped Mott insulators with the strong
short-range antiferromagnetic correlation dominating the entire SC
phase \cite{damascelli03}. In particular, in the the kinetic energy
driven SC mechanism \cite{feng0306}, where the charge carrier-spin
interaction from the kinetic energy term induces a d-wave pairing
state by exchanging spin excitations. Therefore under the kinetic
energy driven SC mechanism, a moderate or strong external magnetic
field aligns the spins of the unpaired electrons, then the d-wave
electron Cooper pairs in cuprate superconductors can not take
advantage of the lower energy offered by a spin-polarized state
\cite{vorontsov10}. In this case, we \cite{huang10} have studied the
magnetic field dependence of the superfluid density and doping
dependence of the upper critical magnetic field in cuprate
superconductors for all the temperature $T\leq T_{c}$ throughout the
SC dome by considering both couplings of the electron charge and
electron magnetic momentum with a moderate and a strong external
magnetic field, respectively, and the results show that the external
magnetic field inducing an reduction of the low-temperature
superfluid density, while the maximal upper critical magnetic field
occurs around the optimal doping, and then decreases in both
underdoped and overdoped regimes, in qualitative agreement with the
corresponding experimental data \cite{sonier99,wen08}. These and the
related results will be presented elsewhere.

\acknowledgments

The authors would like to thank Dr. M. Krzyzosiak and Professor Y.
J. Wang for helpful discussions. This work was supported by the
National Natural Science Foundation of China under Grant No.
10774015, and the funds from the Ministry of Science and Technology
of China under Grant Nos. 2006CB601002 and 2006CB921300.

\begin{widetext}

\appendix
\section{Paramagnetic response kernel at zero-temperature
and superconducting transition temperature} \label{limit1}

In this Appendix, we discuss the paramagnetic part of the response
kernel in Eq. (\ref{kernel5}) in the long wavelength limit at the
zero-temperature ($T=0$) and SC transition temperature ($T=T_{c}$).
Firstly, we discuss the case at $T=0$. In the long wavelength limit,
i.e., $|\vv{q}|\to 0$, the paramagnetic part of the response kernel
(\ref{kernel5}) can be evaluated as,
%\begin{widetext}
\begin{eqnarray}
K_{yy}^{(\rm{p})}(\vv{q}\to 0,0)|_{T\to 0}&=&-2Z^{2}_{\rm{hF}}
{4e^{2}\over \hbar^{2}}\left [{1\over N}\sum\limits_{\vv{k}}
\sin^{2}k_{y} [\chi_{1}t- 2\chi_{2}t'\cos k_{x}]^{2}{\beta
e^{\beta\Ek}\over (e^{\beta\Ek}+1)^{2}}\right ]_{T\to 0}
~~~~\label{kernel6}
\end{eqnarray}
%\end{widetext}
where $N=N_{x}N_{y}$, $N$ is the number of sites on a square
lattice, while $N_{x}$ and $N_{y}$ are corresponding numbers of
sites in the $\hat{x}$ and $\hat{y}$ directions, respectively. In
the d-wave pairing state, the characteristic feature is that the
energy gap vanishes $\Delta_{\rm h}(\vv{k})|_{|k_{x}| =|k_{y}|}=
\Delta_{\rm h}({\rm cos}k_{x}-{\rm cos}k_{y})/2|_{|k_{x}| =|k_{y}|}
=0$ along the diagonal directions in the Brillouin zone. In this
case, the paramagnetic part of the response kernel in Eq.
(\ref{kernel6}) can be rewritten as,
%\begin{widetext}
\begin{eqnarray}
K_{yy}^{(\rm{p})}(\vv{q}\to 0,0)|_{T\to 0} &=&-2Z^{2}_{\rm{hF}}
{4e^{2}\over \hbar^{2}}\left [{1\over N}\sum
\limits_{\vv{k}(|k_{x}|\neq |k_{y}|)} \sin^{2}k_{y}[\chi_{1}t-
2\chi_{2}t'\cos k_{x}]^{2}{\beta
e^{\beta\Ek}\over (e^{\beta\Ek}+1)^{2}}\right ]_{T\to 0}\nonumber\\
&-&2Z^{2}_{\rm{hF}} {4e^{2}\over \hbar^{2}}{1\over N_{x}}\left [
{1\over N_{y}}\sum \limits_{k_{y}}\sin^{2}k_{y}[\chi_{1}t-
2\chi_{2}t'\cos k_{y}]^{2} {\beta e^{\beta\bar{\xi}_{k_{y}}}\over
(e^{\beta\bar{\xi}_{k_{y}}}+1)^{2}}\right ]_{T\to 0}~~~~~
\label{kernel7}
\end{eqnarray}
%\end{widetext}
where $\bar{\xi}_{k_{y}}=Z_{\rm{hF}}(Zt\chi_{1}\cos k_{y}-Zt'
\chi_{2}\cos^{2}k_{y}-\mu)$. The first term of the right-hand side
is equal to zero since the existence of the gap, while the second
term of the right-hand side can be evaluated explicitly as,
\begin{eqnarray}
&-&2Z^{2}_{\rm{hF}} {4e^{2}\over \hbar^{2}}{1\over N_{x}}\left [
\int^{\pi}_{-\pi}{dk_{y}\over 2\pi}\sin^{2}k_{y}[\chi_{1}t-
2\chi_{2} t'\cos k_{y}]^{2}{\beta e^{\beta\bar{\xi}_{k_{y}}}\over
(e^{\beta\bar{\xi}_{k_{y}}}+1)^{2}}\right ]_{T\to 0}\nonumber\\
&=&{2Z_{\rm{hF}}\over Z}{4e^{2}\over \hbar^{2}}{1\over N_{x}}\left [
\int^{\pi}_{-\pi}{dk_{y}\over 2\pi}[\chi_{1}t\cos k_{y}-2\chi_{2}
t'\cos (2k_{y})]{1\over
e^{\beta\bar{\xi}_{k_{y}}}+1} \right ]_{T\to 0}\nonumber\\
&=&{2Z_{\rm{hF}}\over Z}{4e^{2}\over \hbar^{2}}{1\over N_{x}}\left [
\int^{\pi}_{-\pi}{dk_{y}\over 2\pi}[\chi_{1}t\cos k_{y}-2\chi_{2}
t'\cos (2k_{y})]\theta(\bar{\xi}_{k_{y}})\right ],
\end{eqnarray}
which is equal to zero in the thermodynamic limit $N\to\infty$ (then
$N_{x}\to\infty$ and $N_{y}\to\infty$), where the step function
$\theta(x)=1$ for $x<0$ and $\theta(x)=0$ for $x>0$.

Now we turn to discuss the paramagnetic part of the response kernel
(\ref{kernel5}) in the long wavelength limit at $T=T_{c}$
($\beta_{c}=T^{-1}_{c}$). In this case, the energy gap
$\bar{\Delta}_{\rm hZ} ({\bf k})|_{T=T_{c}}=0$, and the paramagnetic
part of the response kernel (\ref{kernel5}) can be evaluated
explicitly as,
\begin{eqnarray}
K_{yy}^{(\rm{p})}(\vv{q}\to 0,0) &=&-2Z^{2}_{\rm{hF}} {4e^{2}\over
\hbar^{2}}{1\over N}\sum\limits_{\vv{k}}\sin^{2}k_{y}[\chi_{1}t-
2\chi_{2}t'\cos k_{x}]^{2}{\beta_{c}e^{\beta_{c}\xik}\over
(e^{\beta_{c}\xik}+1)^{2}}\nonumber\\
&=&-2Z^{2}_{\rm{hF}} {4e^{2}\over \hbar^{2}}\int^{\pi}_{-\pi}{dk_{x}
\over 2\pi}\int^{\pi}_{-\pi}{dk_{y}\over 2\pi}\sin^{2}k_{y}
[\chi_{1}t- 2\chi_{2}t'\cos k_{x}]^{2}{\beta_{c}e^{\beta_{c}\xik}
\over (e^{\beta_{c}\xik}+1)^{2}}\nonumber\\
&=&Z_{\rm{hF}} {4e^{2}\over \hbar^{2}}\int^{\pi}_{-\pi}{dk_{x} \over
2\pi}\int^{\pi}_{-\pi}{dk_{y}\over 2\pi}[\chi_{1}t \cos k_{y}-
2\chi_{2}t'\cos k_{x}\cos k_{y}]{1\over e^{\beta_{c}\xik}+1}
\nonumber\\
&=&Z_{\rm{hF}}{4e^{2}\over\hbar^{2}}{1\over N}\sum\limits_{\vv{k}}
[\chi_{1}t\cos k_{y}-2\chi_{2}t'\cos k_{x}\cos k_{y}]n_{F}(\xik)
={4e^{2}\over\hbar^{2}}[\chi_{1}\phi_{1}t-2\chi_{2}\phi_{2}t']
=-{1\over \lambda^{2}_{L}},
\end{eqnarray}
which exactly cancels the diamagnetic part of the response kernel in
Eq. (\ref{diakernel}), then the Meissner effect is obtained for all
$T\leq T_{c}$ throughout the SC dome.

\end{widetext}


\begin{thebibliography}{00}
\bibitem [*] {add} To whom correspondence should be addressed.
\end{thebibliography}
\end{document}